\documentclass[aps, pre, twocolumn, nofootinbib]{revtex4-1}

\usepackage{graphicx}
\usepackage{dcolumn}
\usepackage{bm}
\usepackage{multirow,caption}
\usepackage{amsmath}
\usepackage{enumitem}
\usepackage{hhline}
\usepackage{soul,color}
\usepackage{xcolor}
\usepackage[normalem]{ulem}
\usepackage{array}
\usepackage{caption}
\usepackage{subcaption}
\usepackage[thinlines]{easytable}

\begin{document}
	\preprint{APS/123-QED}
	
	\title{Evolution of interdependent co-authorship and citation networks}
	
	\author{Chakresh Kr. Singh}
	\email{chakresh.singh@iitgn.ac.in}
	\affiliation{Indian Institute of Technology, Gandhinagar, India-382355}
	\author{Demival Vasques Filho}
	\email{vasquesfilho@ieg-mainz.de}
	\affiliation{Te P\={u}naha Matatini, Department of Physics, University of Auckland, Private Bag 92019, Auckland, New Zealand}
	\affiliation{Leibniz-Institut f\"ur Europ\"aische Geschichte, Alte Universit\"atsstra{\ss}e 19, 55116 Mainz, Germany}
	\author{Shivakumar Jolad}
	\email{shivakumar.jolad@flame.edu.in}
	\affiliation{FLAME University, Pune, India}
	\author{Dion R. J. O'Neale}
	\email{d.oneale@auckland.ac.nz}
	\affiliation{Te P\={u}naha Matatini, Department of Physics, University of Auckland, Private Bag 92019, Auckland, New Zealand}

	\date{\today}
	
	\begin{abstract}
	Studies of bibliographic data suggest a strong correlation between the growth of citation networks and their corresponding co-authorship networks. We explore the interdependence between evolving citation and co-authorship networks focused on the publications, by Indian authors, in American Physical Society journals between 1970 and 2013. We record interactions between each possible pair of authors in two ways: first, by tracing the change in citations they exchanged and, second, by tracing the shortest path between authors in the co-authorship network. We create these data for every year of the period of our analysis. We use probability methods to quantify the correlation between citations and shortest paths, and the effect on the dynamics of the citation-co-authorship system. We find that author pairs who have a co-authorship distance $d \leq 3$ significantly affect each others citations, but that this effect falls off rapidly for longer distances in the co-authorship network. The exchange of citation between pairs with $d=1$ exhibits a sudden increase at the time of first co-authorship events and decays thereafter, indicating an aging effect in collaboration. This suggests that the dynamics of the co-authorship network appear to be driving those of the citation network rather than vice versa. Moreover, the majority of citations received by most authors are due to reciprocal citations from current, or past, co-authors. We conclude that, in order to answer questions on nature and dynamics of scientific collaboration, it is necessary to study both co-authorship and citation network simultaneously. 
		
		
	\end{abstract}
	
	\maketitle
	
	\section{Introduction}
	
	Development of scientific theories and technology is a result of continuous interaction, creation, and effective diffusion of ideas between researchers in the knowledge ecosystem. Digitization of publications and advancements in communication technology have made it easier for researchers to be aware of the existing knowledge capital and possible gaps in the field of study. This facilitation of the spread of scientific knowledge helps researchers to refine their own research methods and to contextualize their work within the domain. It also establishes an indirect interaction between individuals. One might not know a researcher personally but is still aware of her work through technical literature and can gain a sense of familiarity with it. Researchers attend gatherings and conferences to broaden their scope of a subject area and look for new ideas and open problems. Awareness to the state of the art and motivation to solve open problems often becomes a factor in setting up new collaborations between individuals.
	
    Publication of scientific articles provides a narrow, but a well-quantified record of collaboration and exchange of technical information. Interactions between researchers can either be by citing one another or by co-authoring papers together, resulting in a complex system that is changing over time. With such a complex and dynamic system at hand, it is interesting to look for any possible underlying pattern hidden in interactions between researchers, and if these interactions have any mathematical structure. It is possible that such a structure can be used to explain the spread of knowledge and the growth of research fields.
	
	Tools and methods developed within the framework of network science have proven to be very effective in addressing questions of such nature both quantitatively and qualitatively \cite{newman2001structure, newman2001scientific, sinatra2016quantifying}. Effective treatment of complex systems by using networks and easy access to huge databases of publications have attracted a lot of attention in the last two decades. It has also lead to enormous research on structure and evolution of scientific collaboration \cite{dong2017century, tomassini2007empirical}.
	
	Detailed publication records make it easier to create citation networks and co-authorship networks. While the former are directed networks where edges represent citations between nodes, the latter are undirected (either weighted or unweighted) networks where edges exist between two nodes sharing an authorship on a paper. Nodes in these networks can either be papers, authors, universities, and so on, depending on the research question of interest. 
		
    Individual relationships between authors have a great impact on interactions between institutions at meso-level, and countries at macro-level \cite{tomassini2007empirical}, shaping the changing trend of collaboration. Over time, the pattern of collaboration has shown a shift from individual efforts to more cooperative research, increasing the productivity and diversity of scientific publications globally, and resulting in an increase of innovation in this century \cite{dong2017century}. Co-authorship networks have been shown to exhibit small-world characteristics and display high levels of clustering \cite{newman2004coauthorship}. As the network grows and new authors appear, the network structure changes \cite{huang2008collaboration,martin2013coauthorship,vasquesfilho2019structure}. Growing networks also change authors topological position in the network structure, which is found to be directly related to one's productivity and popularity \cite{newman2001structure,newman2001scientific,newman2004coauthorship}. The mechanism for evolving co-authorship networks has shown to exhibit an underlying preferential attachment process \cite{barabasi2002evolution,chen2013evolving}. It is observed that the co-authorship network becomes more connected over time, indicating an increase in collaboration between authors. As a result, the average node distance in the network shrinks. Knowing the co-authorship network structure and its evolution also makes it possible to predict future links between authors by exploiting the changes in an author's neighborhood structure \cite{huang2008collaboration}. Studying co-authorship networks can explain the emergence of new research groups, the significance of some lead researchers, and how one's collaborators change over time.  	
	
	Citation networks, on the other hand, indicate the patterns in generation and diffusion of ideas in the scientific community. Citations received by publications play a significant role in determining their impact as well as their authors' significance in the community \cite{sinatra2016quantifying}.  Evolution of citing patterns can be correlated with the evolution of research fields \cite{gualdi2011tracing,shi2009information}. Since citation networks are directed, tree-like hierarchical structures form the backbone of citations. Patterns in citation dynamics have been extensively explored and modeled. The key feature in citation patterns is the presence of a delay before a paper receives initial citations. Citations acquired by a paper typically increase shortly after publication and reach a maximum within the first few years before decaying with time \cite{pan2018memory,higham2017unraveling}. Considering that the aging effect is important to quantify the probability of getting cited \cite{borner2004simultaneous} or the strength of collaboration in citations and co-authorship networks respectively. The effect of aging can be accounted for in different ways. It can be a weighted measure on edges, that decays since the contributing authors last shared publication \cite{fiala2015ageing} proportional to the time difference between simultaneous participation \cite{tutoky2011time}.  Instead of decaying weights on the edges, adding an aging effect on nodes in the citation networks can also determine the changing probability of receiving citation \cite{hajra2005aging}.
		
	Co-authorship and citation networks together reflect the structure and growth of scientific collaboration. Every new publication results in co-authorship and citation events, therefore it is intuitive that both citation and co-authorship networks complement each other and should have a strong positive correlation in their respective evolution. Many studies addressed these networks and pointed out strong interdependent relations between evolving citations and associated co-authorship networks \cite{kas2012trends,keegan2013structure,martin2013coauthorship,amblard2011temporal,ding2011scientific,tol2011credit,glanzel2004does}. Network measures on time varying graphs for both citation and co-authorship networks exhibit co-dependence of these networks in citing patterns and formation of communities \cite{amblard2011temporal}. Topic modelling algorithms used to assign topics to papers and to investigate citation patterns between authors from similar and different topics fields showed close collaboration between authors working on similar topics. Also, that high profile authors do not generally co-author with one another but do closely cite each other \cite{ding2011scientific}. Large scale data set studies \cite{wallace2012small,martin2013coauthorship} solidify the notion of strong interdependence between citation and co-authorship networks. Citation exchange calculated up to a limited depth of co-authorship connections reveal large gaps in citing patterns of co-authors between natural sciences and social sciences and that the rate of self-citations is constant \cite{wallace2012small}. A detailed analysis of citation and co-authorship networks constructed from a large longitudinal data set (100 years) of publications in Physical Review journals investigated the temporal changes in citing patterns between collaborators \cite{martin2013coauthorship}. One of the main findings of the latter was the constant nature of the fraction of self-citations and citations among co-authors with a strong tendency towards reciprocal citations.
	
	The existing interdependence between the two networks has also helped to define sophisticated weighted measures to distribute the credit of citations between co-authors of a paper, resulting in a more efficient way to calculate authors' significance \cite{tol2011credit}. Studying the citation and co-authorship networks simultaneously not only helps in quantifying a researcher's contribution to the field \cite{gonzalez2015mapping} --- network centrality measures have also proved to be important in quantifying the effect of citation and co-authorship networks on each other \cite{biscaro2014co}. It has been observed that an author's (node's) centrality value in the co-authorship network is a significant factor behind the number of citations received by them \cite{sarigol2014predicting}. Considering the effects of co-authorship networks is also important to define more sophisticated growth mechanisms for citing patterns in networks \cite{guo2017exploiting}. Combined co-authorship and citation networks have been used to predict new collaboration opportunities; that is, new edges in co-authorship networks \cite{lande2016formation} and also to quantify effect of authors and their affiliated institutes international collaborations and region on citations received \cite{sin2011international,yan2012scholarly} by them. Studying both the networks together also helps in forming ranking measures for institutes and researchers \cite{xu2017comes,zhan2017network} in  scientific collaboration.
			
	Simultaneous analyses of citation and co-authorship networks have given insights into  correlations between the two networks. Considering the effect of one network on the other gives a better understanding of the true nature of scientific collaborations. While earlier studies have addressed citation and co-authorship networks simultaneously and established a strong interdependence between the two, there is still scope to understand the details of correlation between these networks. In this study we build on the strong correlation between interacting pairs of authors in citation and co-authorship networks. First, we define a method which tracks the evolution of relationships between each possible author--author pair in both networks. Next, we formulate a null hypothesis for the probability of citation exchange between a pair of authors and use probabilistic analysis to compare it with empirical observation from networks constructed using the publications in the American Physical Society (APS) journals between 1970-2013. This way, we capture both macroscopic and microscopic changes in network structure and address a number of questions which are otherwise difficult to answer.
	
	\begin{enumerate}
		\item What fraction of authors exchange citations but do not co-author however are connected in the co-authorship network?
		\label{item_1}
		\item How are citations exchanged between co-authors? \label{item_2}
		\item How do the statistics in \ref{item_1} vary with network distance between authors?
		\item How does receiving a new citation affect the likelihood of an author creating a new link in the co-authorship or the citation network?
		\item What is the relationship between the probability of citations and network distance?
		\item What is the waiting time distribution for consecutive co-authorship events and for consecutive co-citation events?
	\end{enumerate}

	Our analysis is based on a similar approach used by earlier studies \cite{barabasi2002evolution, biscaro2014co, chen2013evolving, kas2012trends, martin2013coauthorship, yan2012scholarly}. However, our method is significantly different from theirs. By tracing the interactions between all possible pairs of authors in the citation network and at all existing shortest path lengths in the associated co-authorship network, we are able to see in detail the effect of collaborations on citations and vice-versa. Our main contribution is to explain the effect of the distance in the co-authorship network on the citations exchanged between pairs of authors.

	The remainder of this paper is organized as follows: Sec. \ref{sec:methods} presents our methods and explains how we create the distance and citation matrices, as well as how we perform our empirical calculations. In Sec. \ref{sec:results}, we present and discuss our results. We summarize our findings and possible extension of this work in Sec. \ref{sec:conclusion}.
	
	\section{\label{sec:methods}Methods}
	
	For the purpose of our study, we construct a longitudinal data set of publications by Indian researchers in the American Physical Society (APS) journals between 1970 and 2013. Here we consider an author to be Indian if they have any paper with an Indian affiliation. Therefore all papers with authors having at least one Indian affiliation are included in the data set. There were 14,703 such papers \cite{Singh2019}. For the extracted papers we performed name disambiguation on the authors names to assign a unique ID for every author. This was done to account for different naming styles used by authors over time. For naming disambiguation we use edit distance between strings to cluster similar names and then check for neighborhood overlap in the co-authorship network. Names with small edit distance and high neighborhood overlap were grouped together and manually checked for uniqueness using information from online databases. This results in 8,084 unique Indian authors. 	
	
\begin{figure*}[htbp]
		\centering
		\includegraphics[scale = 0.7]{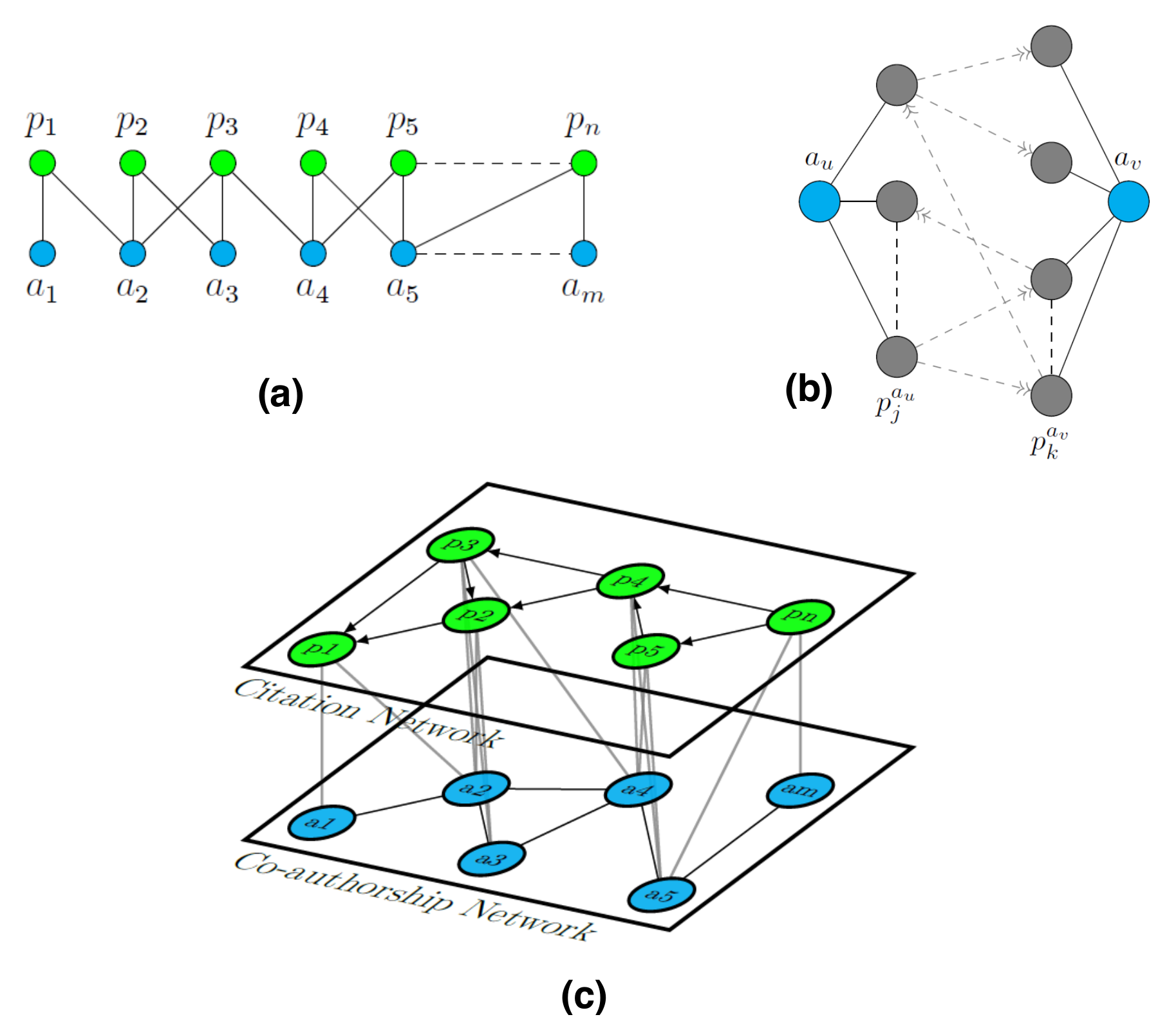}
		
		\captionsetup{singlelinecheck=false, justification=raggedright,  labelsep=space}
		\caption{Diagrams of interdependent citation and co-authorship networks. (a) Bipartite network between papers {$p_i$} and authors {$a_i$} constructed cumulatively. (b) The exchange of citations (dotted arrows) between papers $p_j^{a_u}$ by author $a_u$ and $p_k^{a_v}$ by author $a_v$. (c) A multilayer representation of the interdependence between the citation network and the projected co-authorship network constructed from (a).}
		\label{f1}
	\end{figure*}

	Then, we construct bipartite graphs $B = (U,V,E)$ for every year, from 1970 to 2013, where $V$ is the set of papers, $U$ is the set of authors and $E$ is the set of edges connecting nodes from $U$ and $V$. Each graph $B(t)$ at time $t$ is cumulative, storing all the information until time $t$. From each $B(t)$ we generated weighted, cumulative, and undirected projected co-authorship networks. We also construct cumulative directed citation networks for every year using the paper IDs of Indian publications from the APS citations data set. We illustrate the process of creating theses networks in Fig. \ref{f1}. The ordering of node labels in the projected co-authorship networks is kept consistent for all calculations. Using the above graphs we construct our data matrices for co-authorship and citation networks as follows. 
	
	\subsection{Data Matrices}
	
	In order to aid our analysis, we created two types of matrices, one for the co-authorship networks and the other for the citation networks, for each of the 44 years. The matrices have size $N\times N$, where $N = 8,084$ is the number of unique authors in the whole data set. 
	
	The elements of the first matrix type, $D(t)$, are given by
	
	$d_{ij}(t) =
	\begin{cases} 
	        d  $ if there is a path of length $d$ from $i$ to $j$ $ \\
	        0  $ otherwise$\,,
	\end{cases}$ \\
	such that the matrix $D(t)$ captures the distances between all possible  $\binom{N}{2}$ pairs of authors in the network at time $t$. For $d_{ij} = 0$, it can mean that the nodes do not exist in the network at that time or that they are not connected via any path.
	
	The second matrix type, $C(t)$, stores the citations exchanged between papers written by $i$ and $j$ \textit{until} a given time $t$. That is, $c_{i \leftarrow j}(t)$  is the cumulative number of times that $j$ cites $i$ until that particular year.
	
\begin{figure}[htbp]  
		\centering
		\includegraphics[scale = 0.29]{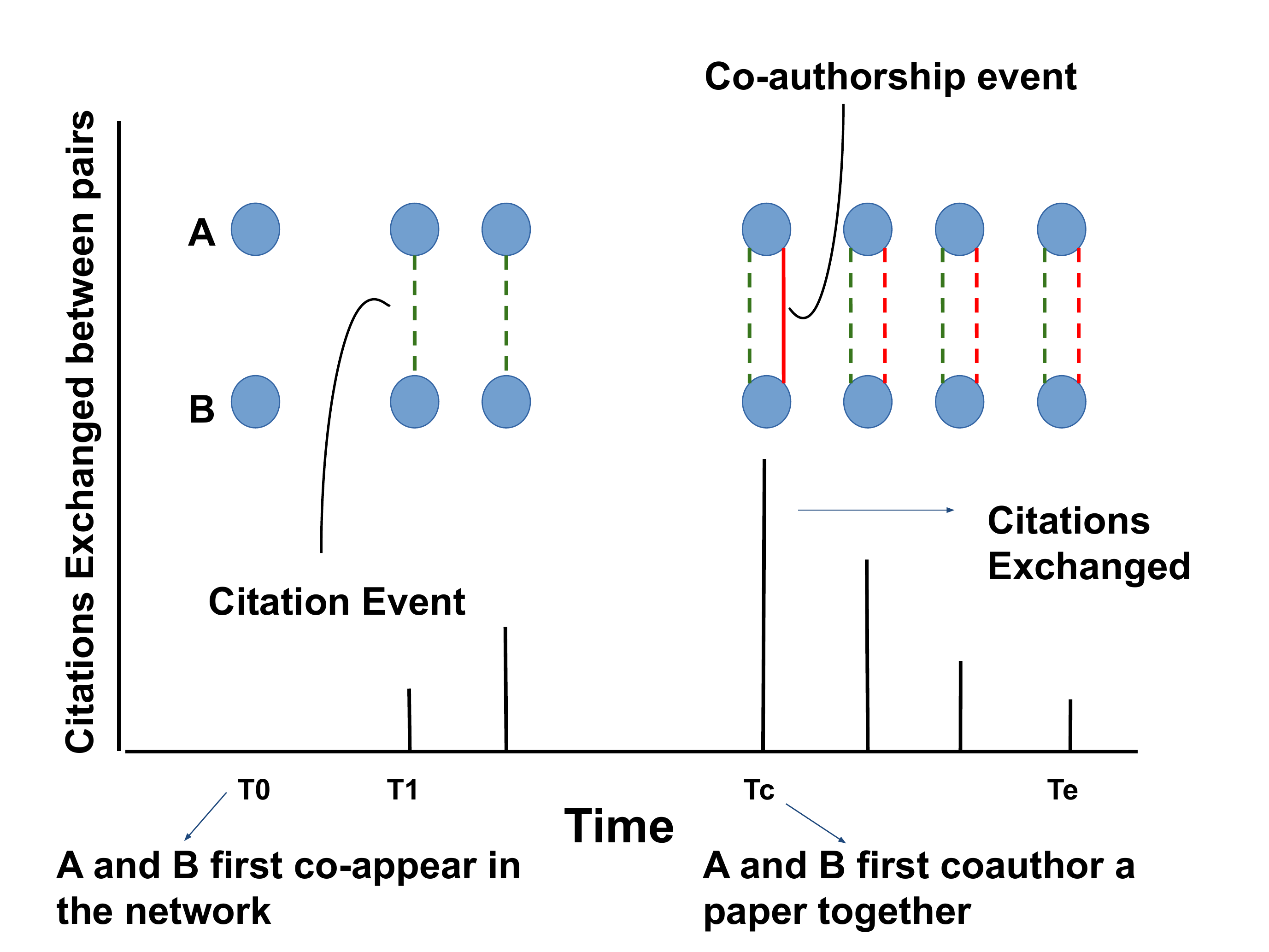}
		
		\captionsetup{singlelinecheck=false, justification=raggedright,  labelsep=space}
		\caption{Schematic diagram of the method of counting the number of citations exchanged between pairs of authors before and after their first collaboration. Here we are interested to see the difference in mutual citations before and after $T_C$. Author pairs that have $T_0 = T_C$ are ignored here.}
		\label{f2}
	\end{figure}

	To see the aging effect in collaboration between authors, we trace the history of co-authorship events. First, we extract the distance $d_{ij}=1$ collaborations at the end of the time period (i.e. edges in the 2013 co-authorship graph). Next, we trace the presence of edge between $i$ and $j$ (contributing authors) in reverse order. The time when the edge first appears is marked $T_c$, which is the year of first collaboration. Then, we check for the presence of $i$ and $j$ in the network before $T_c$ until we find $T_0$, the first year in which authors $i$ and $j$ both are present in the network. For all these years, the number of citations exchanged by pairs of authors, at every time step before and after collaboration, are recorded. A diagram illustrates this method of tracing the history of collaboration between pair of nodes in Fig. \ref{f2}. 
	
	In order to compare the history of citation and co-authorship for all pair of authors, we shift the time series on the x-axis of every pair to zero. That is, we adjust $T_c$, the year of first collaboration between two authors, to 0. Out of these, we remove the pairs that had $T_0 = T_c$. We do this to remove authors who appear in the network together with a shared publication. As these authors will not have any citing history prior to their first collaboration, we exclude them. The remaining are pairs of authors who took at least one year to collaborate after appearing together in the network.
	
	\subsection{Calculations}
	The data matrices $D(t)$ and $C(t)$ store the information of the distance and the citations exchanged between all possible pairs of nodes for the co-authorship and citation networks, respectively, for every time step (which is an year in our case). This enables us to calculate any changes in distance or citations exchanged from one year to another. Then, to address the research questions mentioned in the Introduction, we define our calculations based on the different situations that each possible pair of nodes present in the networks. More specifically, we calculate the following, at every time step $t$:
	
	\begin{enumerate}
	    \item What fraction of authors exchange citations but do not co-author?
	
	    \begin{enumerate}[label=(\alph*)]
    
	        \item We count pairs that exchange citations ($(c_{ij} + c_{ji} \neq 0$) and have a connected path in the co-authorship network ($d_{ij} > 1$).
	
	        \item We count pairs that exchange citations ($(c_{ij} + c_{ji} \neq 0$) but are not connected in the co-authorship network $d_{ij} = 0$. These are the pairs that are aware of each others work via citations, but do not have any direct or indirect connection with each other via co-authorship.  
	    
	    \end{enumerate}
	
	    \item How are citations exchanged between co-authors?
	    
	    \begin{enumerate}[label=(\alph*)]
	    
	        \item We count pairs that co-author ($d_{ij} = 1$)  but do not exchange citations $c_{ij} + c_{ji}=0$.
	        
	        \item We count pairs who are co-authors ($d_{ij} = 1$) and exchange citations ($c_{ij} + c_{ji} \neq 0$).
	   
	   \end{enumerate}
	   
	   \item How do the statistics in 1 vary with network distances between authors? We count pairs that exchange citations ($c_{ij} + c_{ji} \neq 0$) for different distances $d_{ij}$.  
	    
	   \item How does receiving a new citation affect the likelihood of an author creating a new link in the co-authorship or the citation network?
	   
	   \begin{enumerate}[label=(\alph*)]
	   
	        \item \textbf{Response by authors, in terms of citations (how they cite back), to other authors who cited them, and to the total citations they received.} For every author $i$ in the citation matrix $C(t)$, we define $N_i$ as the number of authors that cite $i$. Among $N_i$, $n_i$ is the number of authors whom $i$ cites back. Similarly, $C_{in}^{N_i}$ are the total citations received by $i$ from the set $N_i$ of authors and $C_{out}^{n_i}$ are the total number of citation given out by $i$ to the set of $n_i$ authors. The response by an author is calculated as: a) $n_i/N_i$ --- the response to citing authors; and b) $C_{out}^{n_i}/C_{in}^{N_i}$ --- the response to citations received. 
	   
	   \end{enumerate}
	   
	   \item What is the relationship between the probability of citations and network distance ?
	   
	   \begin{enumerate}[label=(\alph*)]
	   
	    \item \textbf{Correlation between citations exchanged and network distance.} In order to define the probability of citation between author, we take ${\cal P}_{ij}^{C}(t)$ as the probability that $j$ cites $i$ at time $t$, given by
	    \begin{equation}
	    \label{eq:pij_ct}
	        {\cal P}_{ij}^{C}(t) = f_i^c(t)c^j(t^*) \,, 
	    \end{equation}
	    where $f_c^i(t)$ is the fraction of citation $i$ has received prior to $t$, and $c^j(t^*)$ is the fraction of citations $j$ gives out in the year between $t-1$ and $t$. 
	    We also define the mean probability
	    \begin{equation}
	        {\cal P}^{C}(t) = \langle {\cal P}_{ij}^{C}(t)\rangle \,.
	    \end{equation}
	
	    \end{enumerate}
	
	\end{enumerate}
	
	We calculate $f_c^i(t)$ and $c^j(t^*)$ from our data matrices according to
	
	\begin{equation}
	f^c_i(t) = \frac{\sum_j c_{ij}(t)}{\sum_i \sum_j c_{ij}(t)} \,,
	\end{equation} 
	\begin{equation}
	c^j(t^*) = \frac{\sum_i c_{ij}(t^*)}{\sum_i \sum_j c_{ij}(t^*)} \,,
	\end{equation}
	where
	\begin{equation}
	c_{ij}(t^*) = c_{ij}(t) - c_{ij}(t-1).
	\end{equation}
	
    Our reasons behind this approach are twofold: (i) Popular authors (or papers) have a greater tendency to get cited ($f_c^i(t)$) --- the frequently observed preferential attachment phenomenon; and (ii) if an author (paper) is giving out more citations it uniformly increases the probability of other authors (papers) getting cited ($c^j(t^*)$). It should be noted that this definition is independent of the relationship between authors in the co-authorship network. This will serve as the null model for our subsequent comparisons since it calculates probability distributions for citations without accounting for co-authorship distance. 
	
	First, Eq. (\ref{eq:pij_ct}) estimates the probability of a directed edge between authors in the citation graph, given no other information than the number of citations exchanged between authors recorded in $C(t)$. Next, we need to define the relations to compare the empirical observations from the citation and co-authorship networks with the null model. We use Bayes' theorem to construct probability relationships. This helps us to put constraints in our observations that will help us to highlight the dependency of citations on the shortest path between authors in the co-authorship network. 
	  
	Then, if $T_0$ is the time of first co-appearance of $i$ and $j$ and $P(d=1)$ is the probability of them co-authoring, we have
	\begin{equation}
	P(d=1|C^{>0}(t)) = \frac{P(C^{>0}(t)|d=1)\times P(d=1,t)}{P(C^{>0}(t))} \,,
	\label{eq_6}
	\end{equation}
	where 
	\begin{equation}
	P(C^{>0}(t)) = \sum_{k=0}^{\infty} P(C^{>0}(t)|d=k) \times P(d=k)
	\label{eq_7}
	\end{equation}
	
	Eq. (\ref{eq_6}) is the empirical probability of observing pairs of authors connected by a path of length one, given that they exchange non-zero citations ($C^{>0}(t)$), normalized by the probability of non-zero citations for pairs at all possible shortest path lengths (Eq. (\ref{eq_7})). Therefore for any distance $d=k$, we have
		
	\begin{equation}
	P(d=k|C^{>0}(t)) = \frac{P(C^{>0}(t)|d=k).P(d=k)}{P(C^{>0}(t))}
	\end{equation}
	
	If we reverse the relationship using Bayes' rule, the empirical probability of observing pairs with non-zero citations between them, given they are at distance $d=k$ is calculated according to 
    
   	\begin{equation}
   	    \begin{split}
            P(C^{>0}(t)|d=k)  = \hspace{6cm}\\ \frac{P(d=k|C^{>0}(t)) \times P(C^{>0}(t))}{P(d=k|C^{>0}(t))P(C^{>0}(t)) +P(d=k|C^{0}(t))P(C^{0}(t))}    
   	    \end{split}
	\label{eq_9}
	\end{equation}
	
	The denominator in Eq. (\ref{eq_9}) normalizes over pairs that either exchange citations ($C^{>0}(t)$) cite or do not cite each other ($C^{0}(t)$) given a network distance $d=k$.
	
	\begin{enumerate}[resume]
	    \item What is the waiting time distribution for consecutive co-authorship events and for consecutive co-citation events? 
	    
	    \begin{enumerate}[label=(\alph*)]
	    
	        \item \textbf{Co-authorship events.} The co-authorship networks are weighted undirected networks constructed cumulatively. Therefore, every time an author shares a paper with another one, the weight of the edge between them in the co-authorship network changes. We record the time $\Delta t$ it takes for this change to happen. We do so for all pairs of authors in the network over time.
	        
	        \item \textbf{Co-citation events.} For every pair $ij$ in the citation matrix $C(t)$ we record the time $\Delta t$ it takes for a change in the value of $(c_{ij}+ c_{ji})$ over the whole time period. 
	        
	    \end{enumerate}
	    
	\end{enumerate}
	
	With our data matrices, $C(t)$ and $D(t)$, and the distinct relations between citations exchanged and distance in the co-authorship networks defined, we now turn our focus to the understanding of the interdependence between the co-authorship and the citation networks. That is what we address in the next section.
	
	\section{\label{sec:results}Results and Discussions}
	
\begin{figure*}
		\centering
		\includegraphics[scale = 0.65]{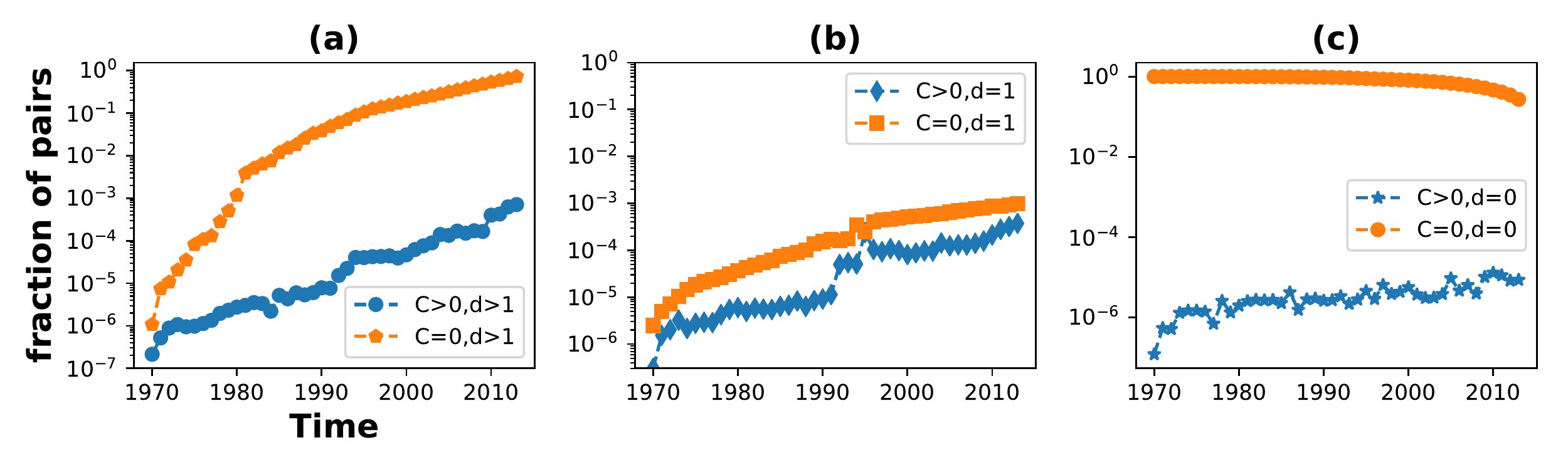}
		\captionsetup{singlelinecheck=false, justification=raggedright,  labelsep=space}
		\caption{Number of pairs of authors with different citation and distance relations changing with time. Each sub-figure is for pairs at different shortest path length in the co-authorship networks. (a) Pairs that are connected with path length greater than one, (b) pairs that co-author a paper, and (c) pairs that are not connected at all. The orange line in each represents pairs that do not exchange any citations while, the blue one is for pairs that exchange at-least one citation. The number of pairs at each time are divided by total possible pairs. I.e. $\binom{N}{2}$ where N=8,084. The blue line in each figure suggests that a very small fraction of the total possible pairs in the co-authorship network are responsible for all the citations observed. We are interested in seeing the patterns in citations between these pairs.}
		\label{f3}
	\end{figure*}

	Using the data matrices $C$ and $D$, described above, we count the number of pairs for different citation and co-authorship distance relations as the networks evolve with time.  To understand the interdependence between the citation network and its associated co-authorship network we calculate the citations exchanged between pairs by splitting them into three groups:
	\begin{enumerate}
	    \item  $D_{ij}>1$: Pairs that are connected with a shortest path length $d>1$ in the co-authorship network (Fig. \ref{f3}(a)), 
	    \item $D_{ij} = 1$: Pairs that co-author a paper together (Fig. \ref{f3}(b)), and  
	    \item $D_{ij} = 0$: Pairs that are not connected in the co-authorship network (Fig. \ref{f3}(c)).
	\end{enumerate}

	For each group we separately count the number of pairs of authors that do not exchange any citation (orange line in Fig. \ref{f3}) and pairs that have non-zero citations shared between themselves (blue line in Fig. \ref{f3}). For all cases, as the citation and co-authorship networks grow, the fraction of pairs that do not have any citations between them is larger than the the fraction of pairs that do exchange citations. 
	
\begin{figure}
	    \centering
		\includegraphics[scale = 0.49]{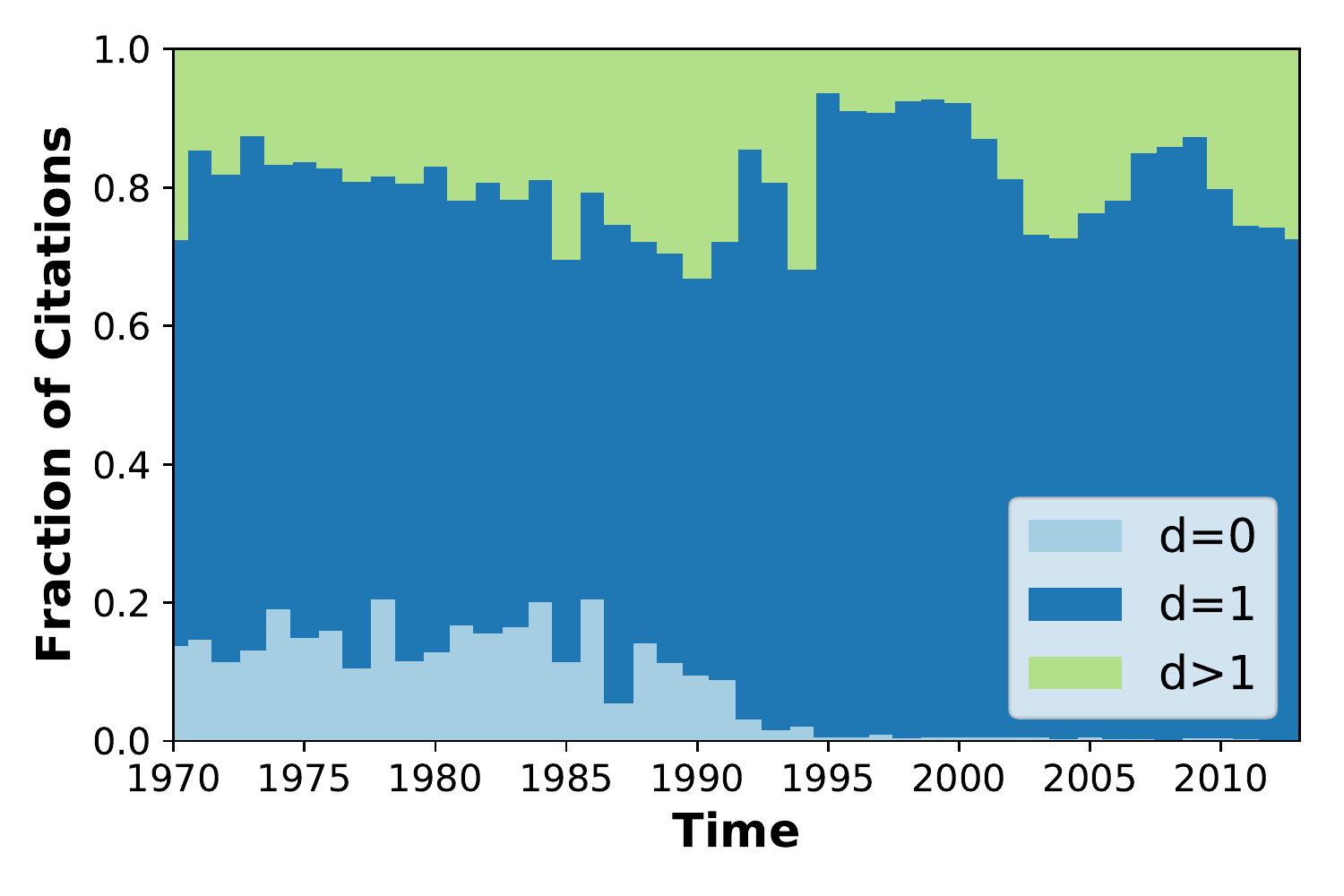}
		\captionsetup{singlelinecheck=false, justification=raggedright,  labelsep=space}
		\caption{Above figure represents the fraction of citations exchanged between pairs at different network distances over time distributed between pairs at different path length as in (a), (b) and (c) in Fig.\ref{f3}. The figure suggests that the majority of citations are exchanged between co-authors (blue region) $d=1$. Distant connections $d>1$ (green region) also contribute to the total citations however disconnected pairs $d=0$ (sky blue region) contribute in the beginning with almost negligible effect in later time to the total citations.}
		\label{f4}
	\end{figure}

	The contributions of each of the three groups defined to the total number of citations are shown in Fig. \ref{f4}. First, the light green region in Fig. \ref{f4} is the fraction of citations exchanged between pairs that have distance $d>1$ in the co-authorship networks. Even though the number of such pairs is a small fraction in the co-authorship networks (blue line in Fig. \ref{f3} (a)) they still contribute significantly to the total number of citations. Second, pairs that co-author are responsible for most of the citations (blue region in Fig. \ref{f4}), which reflects the importance of an authors' collaborators to the number of citations received. And third, disconnected pairs exchange a very small fraction of total citations between them (sky blue region in Fig. \ref{f4}), showing a decreasing trend, until it almost vanishes in the last years. 
	
	The behavior described above could be a consequence of our choice of the data set. Since we focus on a small fraction of the total number of publications (those from Indian authors) in the global APS network, authors are expected to be well connected and closely citing each other. As most of the pairs are connected by a path in a growing co-authorship network --- represented by the orange line in Fig. \ref{f3}(a) approaching one --- and are aware of each other in the network (the decrease in the orange line in Fig. \ref{f3}(c)) the citations by distant pairs decreases. The trend in the number of co-author pairs that exchange citations shows an interesting sudden jump in the mid 1990s. We believe this trend is due to the increasing number of nodes in the co-authorship network. In the beginning there were very few nodes (authors) in the network, most of whom appeared as co-author pairs which explains the initial increase. Post 1993 (the blue line in Fig. \ref{f3}(b)) we notice a sudden increase in the number of such pairs. This sudden change is because of the introduction of papers with a high number of authors in that period. These papers lead to large cliques (totally connected sub-graphs) in the co-authorship network. Most of these papers are published by large collaboration groups often having multiple common authors in their publications and with many Indian authors being part of such groups. For example, the papers \cite{abachi1994first, abachi1995top, abelev2010identified} have 351, 395, and 383 authors respectively. Even one citation shared between such papers would dramatically inflate the number of citations exchanged, due to the large size of the induced co-author cliques.  
	 
\begin{figure}
		\centering
		\includegraphics[scale = 0.49]{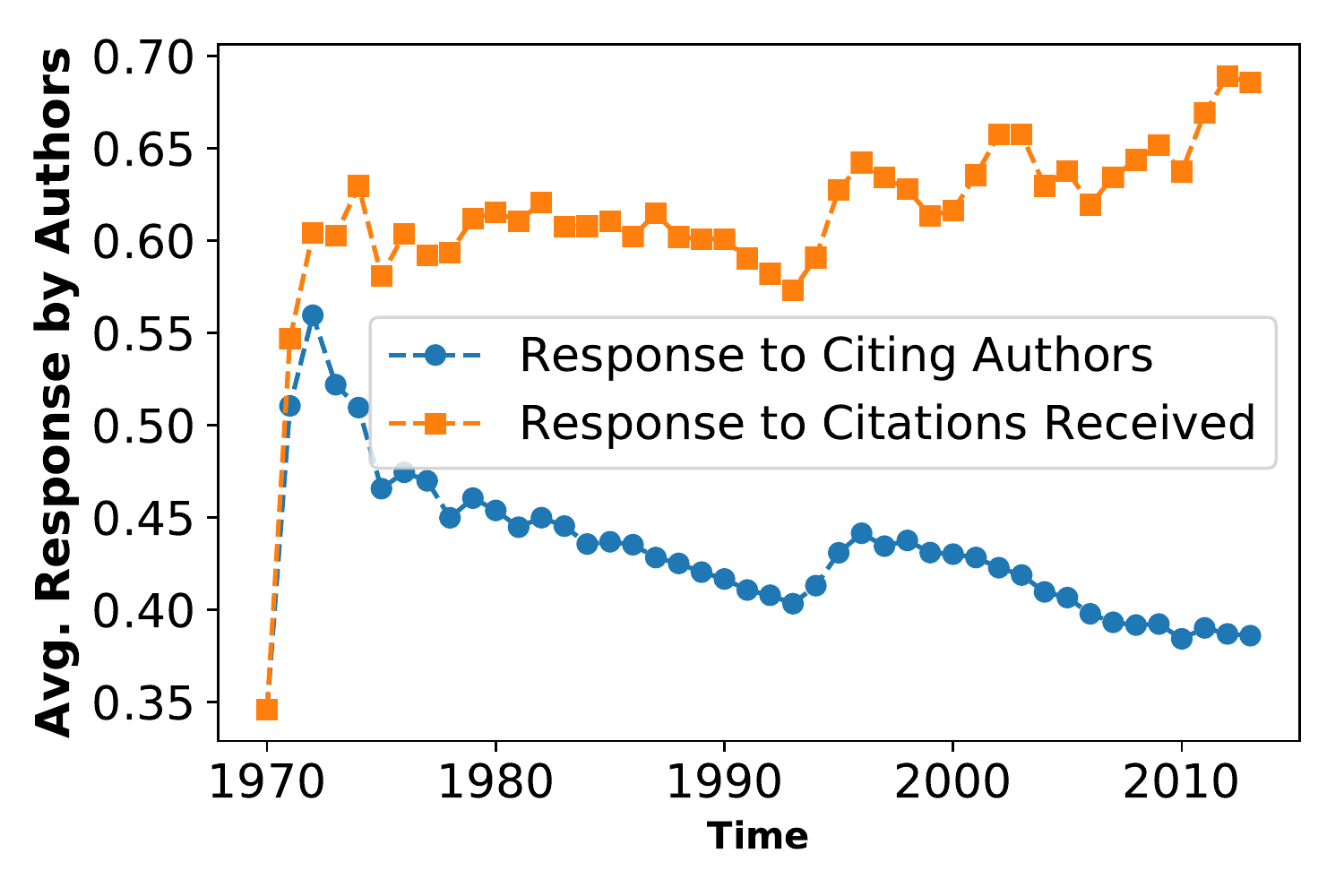}
		\captionsetup{singlelinecheck=false, justification=raggedright,  labelsep=space}
		\caption{The increasing mean probability for authors to give back a citation for every citation they receive (orange squares) indicates the tendency of authors to cite more, i.e. longer reference list. On the other hand, the decreasing mean probability for authors to cite back someone who cited them in the past shows the aging effect in mutual citations.}
		\label{f5}
	\end{figure}

	In the above calculations, we counted the total citations exchanged between the pairs in the co-authorship network for different network distances normalized by the total number of possible pairs in the co-authorship network. The citation count included both incoming and outgoing citations. To measure the response of authors to incoming citations we split our calculations in two parts. First, we calculate the average fraction of out-going citations from authors for every citation received by them. We observe that over time people tend to cite more and more articles in their work hence, we see an increasing trend in response to citations (the orange curve in Fig. \ref{f5}). Second, we calculate the average fraction of incoming citations that an author responds to by subsequently citing the author who initially cited her (the blue curve in Fig. \ref{f5}). When the co-authorship network is in its initial phase, with a small number of researchers,  most co-author pairs cite each other. Hence, we observe high citation reciprocity at initial times. As the network grows, the distribution of citations becomes more heterogeneous as some authors receive more citations than others (authors of influential papers receive many citations). In addition, authors that are no longer publishing cannot reciprocate anymore but still receive citations. Thus, more citations are given out than received in average, which results in a decreasing trend in citation reciprocity (the blue line in Fig. \ref{f5}). The sudden increase in reciprocity in the mid 1990s is due to the citations exchanged between papers with a large number of authors, which, as aforementioned, started to appear around that time.
		
	So far, our observations give a macroscopic understanding of the interdependence of simultaneously growing citation and co-authorship networks. To probe that further, we make more elaborate calculations to see the effect of co-authorship network distance on the citations exchanged. In the interest of this objective we ignore pairs that do not exchange any citations (orange line in Fig. \ref{f3}) from our subsequent analysis. 
	
\begin{figure}
		\centering
		\includegraphics[scale = 0.49]{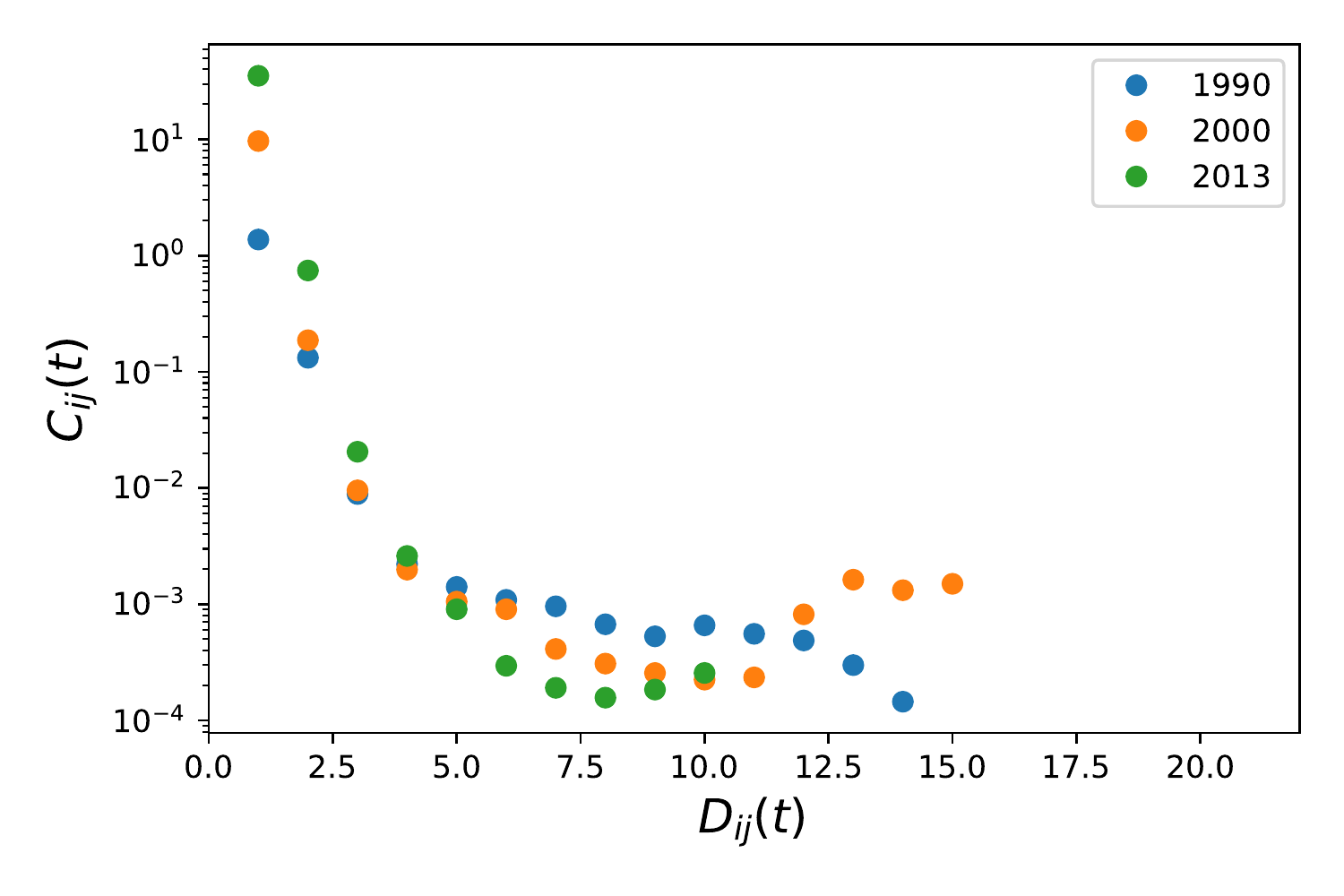}
		\captionsetup{singlelinecheck=false, justification=raggedright, labelsep=space}
		\caption{Average total number of citations between pairs of authors vs distance $d$ between pairs in the co-authorship network (for years till 1990, 2000 and 2013). Consistency in the trend for every time period indicates an interdependence between mutual citations between pairs and their co-authorship network distance. Average citations fall rapidly up to $d\leq3$ and have a similar trend for longer distances.}
		\label{f6}
	\end{figure}

\begin{figure}
		\centering
		\includegraphics[scale = 0.5]{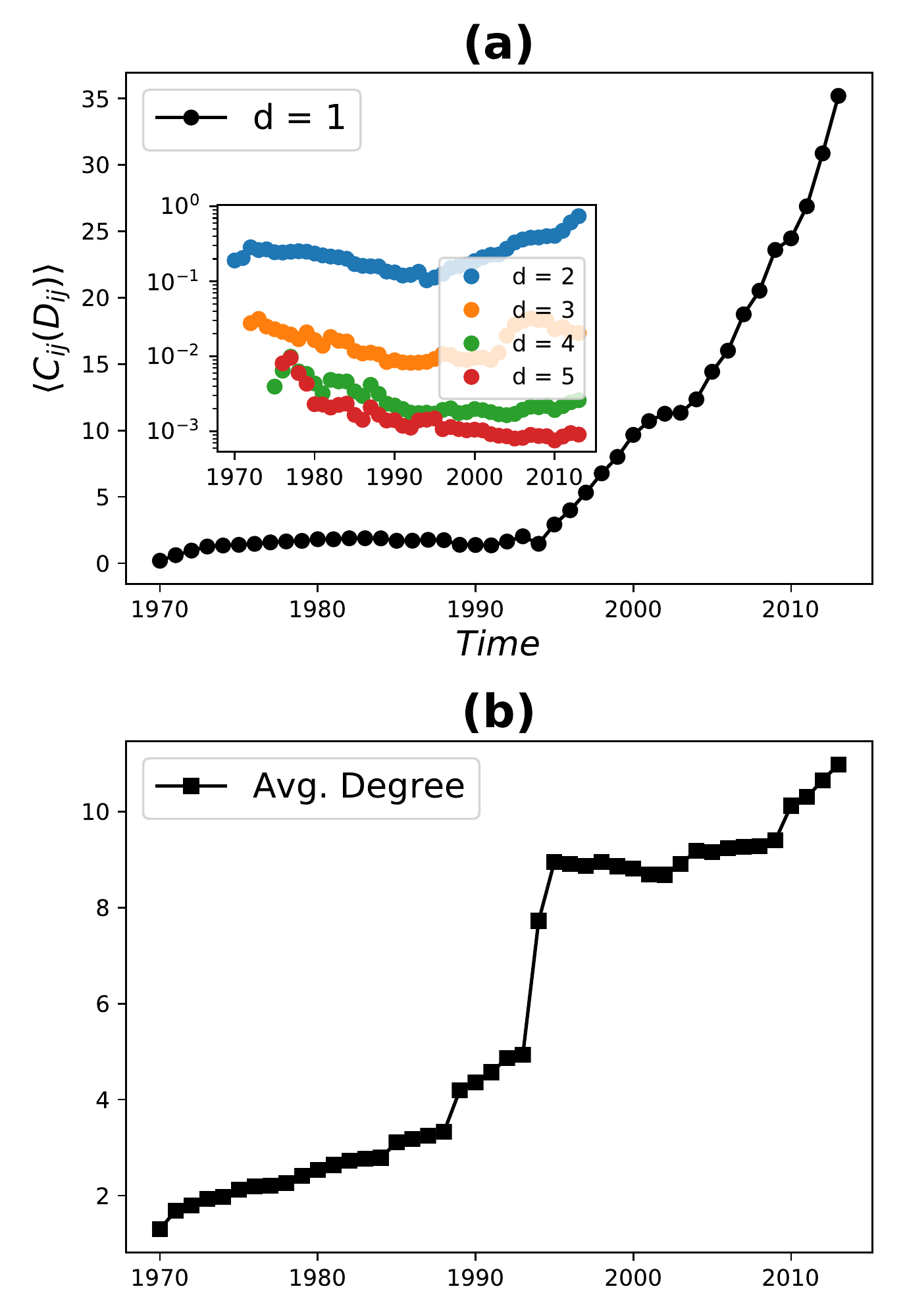}
		\captionsetup{singlelinecheck=false, justification=raggedright, labelsep=space}
		\caption{ (a) Change in the average number of citations exchanged between pair $i,j$ at distance $d=1$ with time. The sudden increase in citations for pairs at $d=1$ post 1994 can be attributed to sudden increase in the degree in (b) i.e. number of collaborators. The inset in (a) shows the variation of citations for higher network distances. We see a significant difference (almost by an order of magnitude) between $d = 1,2 \textrm{ and } 3$. For $d>3$ the trend is similar. Here we have plotted only up to $d=5$ for clarity.}
		\label{f7}
	\end{figure}

    The number of citations exchanged by pairs of authors $ij$ at distance $d_{ij}$ in the co-authorship network decreases rapidly with increasing co-authorship distance (Fig. \ref{f6}). We plot this relation for networks at different times (1990, 2000, and 2013) to show that the trend is consistent as the network evolves. The average citations between pairs (Fig. \ref{f7}(a)) displays significant difference in the temporal trends for different network distances. For direct collaborations (pairs with $d_{ij}=1$), the rate of citation exchange increases with time, more rapidly after 1995. This is likely due to a sudden increase in the number of collaborators (as seen by the change in the average degree of nodes in Fig. \ref{f7}(b)). The average number of citations differ roughly by an order of magnitude for co-authorship distances of $d_{ij}=1, 2 \& 3$; for larger distances ($d\geq 4$), the average number of citations exchanged are very low, with similar trends, as show in Figs. \ref{f7}(a) and \ref{f7}(a) (inset). 
	
\begin{figure}[htbp]  
		\centering
		\includegraphics[scale = 1]{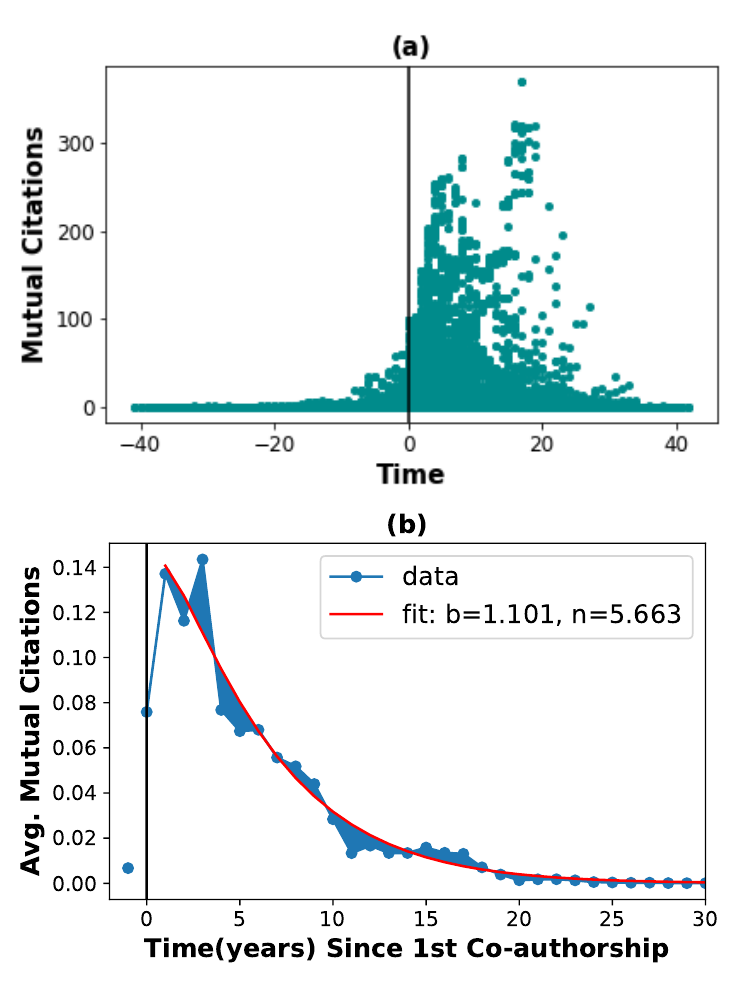}
		\captionsetup{singlelinecheck=false, justification=raggedright,  labelsep=space}
		\caption{(a) Scatter plot of citations exchanged before and after time of first co-authorship event (black vertical line) between authors. Authors exchange more citations immediately after they co-author a paper together. This mutual citation then decreases with time. (b) The average number of citations at each time (measured relative to the time of initial co-authorship) fitted to a Weibull distribution. The trend indicates an aging effect in mutual citation between pairs of authors. }
		\label{f8}
	\end{figure}

	The change in the strength of collaboration between two authors $i$ and $j$ is measured by the total citations exchanged between $i$ and $j$ (scatter plot Fig. \ref{f8}(a)) before and after the period $T_c$ of first collaboration (black line in Fig. \ref{f8}(a)). This includes cases where $i$ and $j$ cite their own previous papers. Co-author pairs exhibit an interesting citing pattern --- the number of citations shows a steep rise after the first co-authorship event and then decays with time, indicating an aging effect.
	
	Interestingly, the peak for this distribution is within five years of $T_c$. When averaged over all times, we notice that the decaying trend is well fitted by a Weibull distribution, $f(t) = \frac{b}{n}(\frac{t}{n})^{b-1} e^{-(\frac{t}{n})^b}$,	as was noted in \cite{borner2004simultaneous}.
	
\begin{figure}
		\centering
		\includegraphics[scale = 0.49]{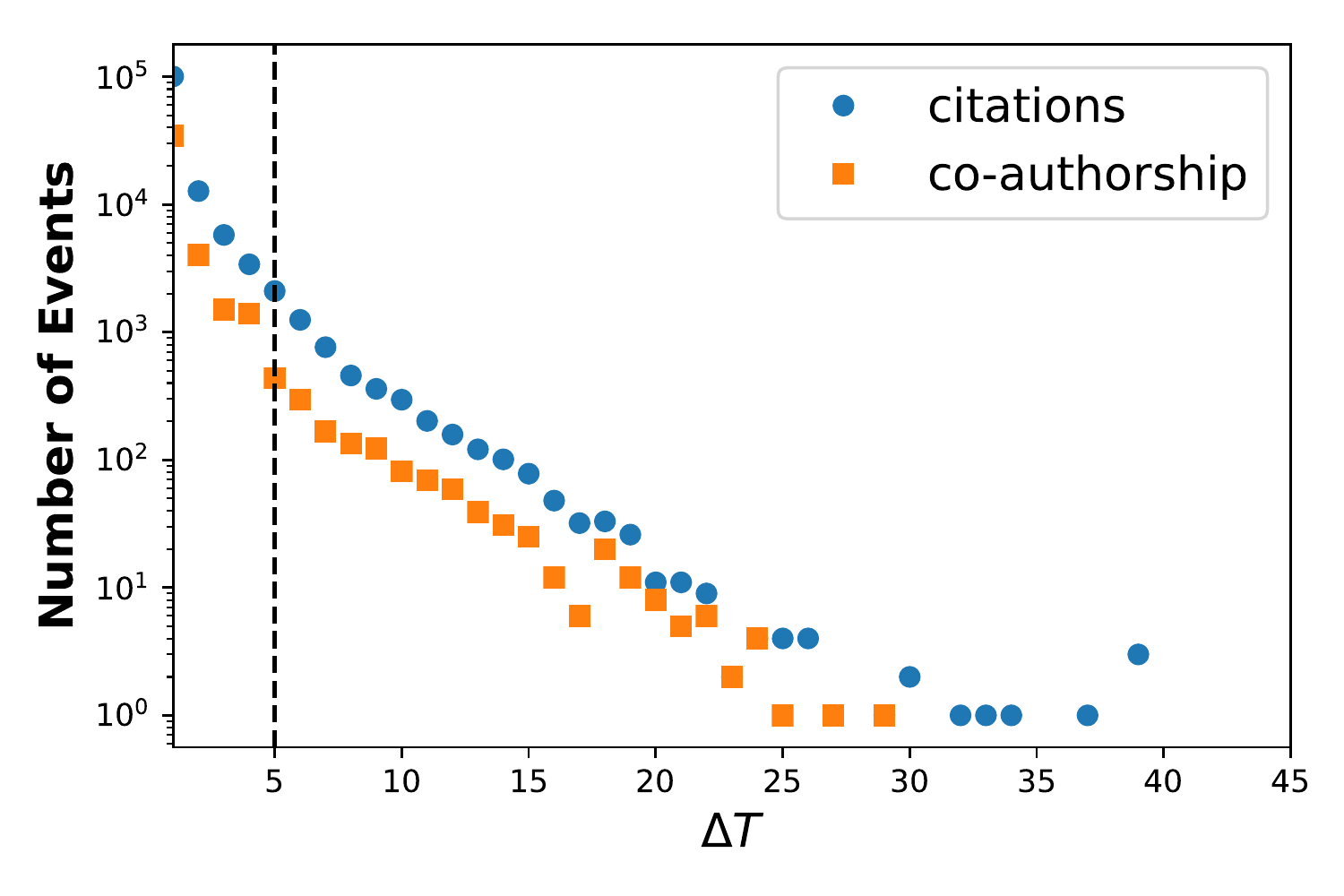}
		\captionsetup{singlelinecheck=false, justification=raggedright,  labelsep=space}
		\caption{Waiting time distributions between consecutive citations and consecutive co-authorship events for  pairs of authors. the black dashed line indicates the five year mark, before which more than $95\%$ percent of citation and co-authorship events occur.}
		\label{f9}
	\end{figure}

	Next, we calculate the waiting time ($\Delta T$) distribution for consecutive citations between pairs of authors and for consecutive co-authorship events. Both follow a similar trend (Fig. \ref{f9}) with the majority of co-authorship and citation events ($95\%$) happening within the first five years of the initial event (black dotted line in Fig. \ref{f9}).
	        
\begin{figure}
		\centering
		\includegraphics[scale = 0.49]{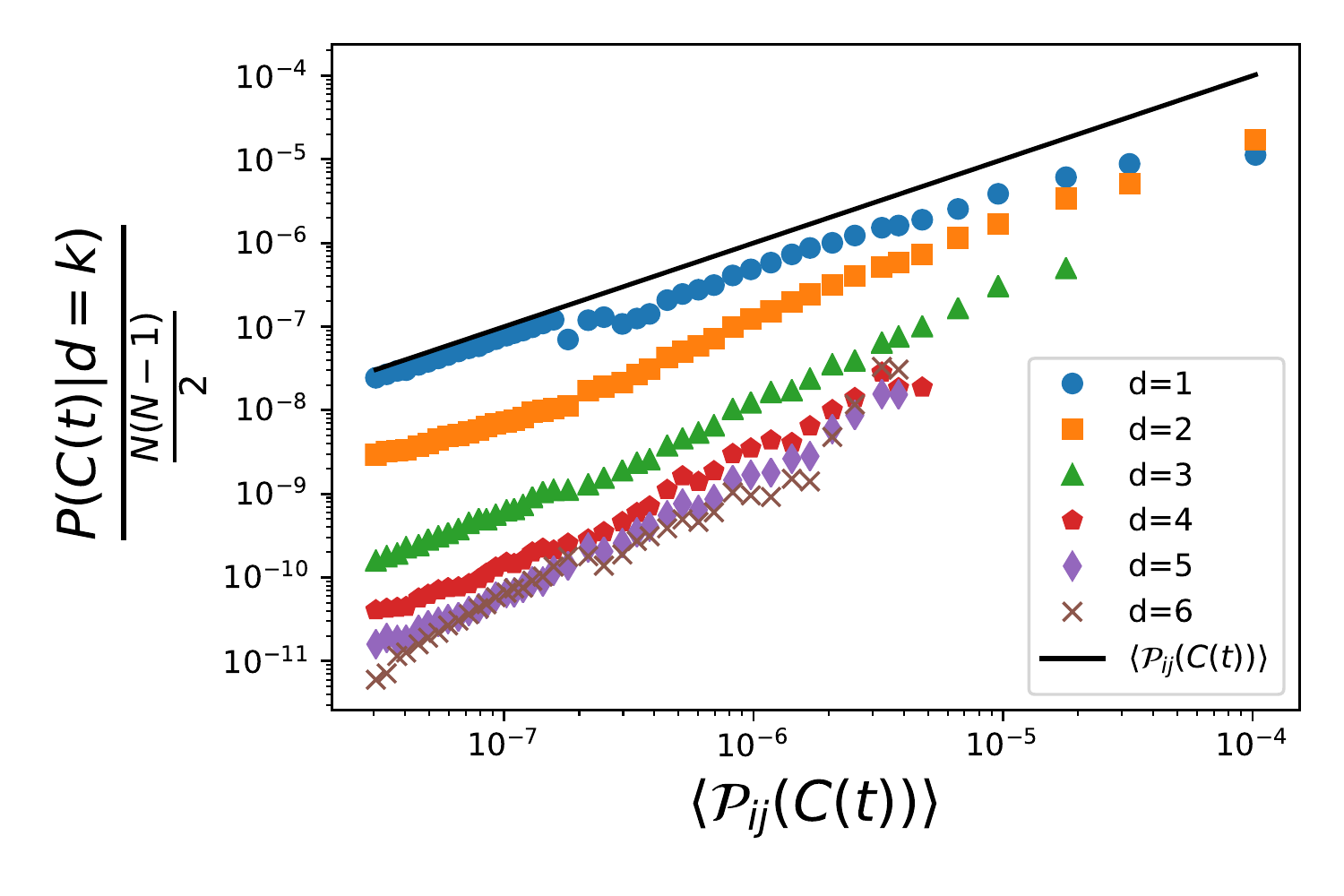}
		
		\captionsetup{singlelinecheck=false, justification=raggedright,  labelsep=space}
		\caption{Comparing the probability obtained from Bayes' formalism vs the hypothesis for different network distances. A clear difference in probabilities for different distances between pairs indicates a distance dependence in citation between authors.}
		\label{f10}
	\end{figure}

	Finally, we calculate the empirical probabilities of authors acquiring citations, firstly with the null model (Eq. (\ref{eq:pij_ct})) and subsequently with the empirical probabilities derived using Bayes' formalism (Eqs. (6-9)). The latter accounts for co-authorship distance in determining the probability of citations exchanged between pairs (Fig. \ref{f10}). We notice that the null model, which is proportional to popularity of the author (paper), and the number of citations given out by the citing author (paper) are not sufficient to explain the observed behavior of citation exchange. The citing patterns significantly differ for different network distances between author pairs. The probability of citations between pairs at $d_{ij}=1$ (blue line in Fig. \ref{f10}) closely follows the null model, while pairs at greater distances significantly differ from it. From this we infer that most of the overall citation behaviour is explained by only considering those citations that come directly from co-authors with $d_{ij}=1$. This is also made evident by the blue region in Fig. \ref{f4} where citations from co-authors contribute to most of the total number of citations. Distinct probabilities of citations at different co-authorship distances (Fig. \ref{f10}) and the decay in average citations with higher network distances (Fig. \ref{f6}) indicate an interdependence between citing patterns and co-authorship network distances, hence confirming our hypothesis. 
	
	By splitting our analysis into different research questions, we were able to explore both macroscopic and microscopic trends in citing patterns between authors as they appear in the associated co-authorship network.
	
	In our macroscopic approach we first count connected pairs, co-authors and disconnected pairs in Fig. \ref{f3} and their contributions to total citations exchanged in Fig. \ref{f4} to find that 
	\begin{itemize}
	    \item Very small fraction of pairs are connected with distance $d_{ij} > 1$ in the co-authorship network but still have a significant contribution to overall number of citations.
	    \item Co-authors are a very small fraction of the total possible number of pairs but account for most of the citation exchanges observed.
	    \item Disconnected pairs contribute to citations in the beginning of the network, but their contribution becomes negligible as the network grows and becomes more connected. 
	\end{itemize}
	
	Besides interactions between pairs, we also investigated the average probability of an author citing back an author that has cited her (the blue line in Fig. \ref{f5}) and the average probability of an author giving back a citation for every citation received (the orange line in Fig. \ref{f5}). The trends indicate that while, over time, the ratio of outgoing to incoming citations per author has increased. However, the  ratio of pairwise outgoing citations that reciprocate citing authors decreases over time. In Fig. \ref{f8}, the plot of citations exchanged between authors exhibits a sudden increase when they co-author a paper and then decays, which is consistent with the aging effect in citations and collaboration reported by earlier studies. A similar effect is observed in Fig. \ref{f9} where $95\%$ of consecutive co-authorship and citation events happen within the first five years of an initial co-authorship event. 
	
	On the other hand, in microscopic calculations, we first calculate the average citations shared between author pairs at all possible network distances for networks at different points in time (Fig. \ref{f6}) and for all time steps (Fig. \ref{f7} --- plotted only up to $d=5$ for clarity). The number of average citations shows a steep decay up to $d\leq3$ and then is almost stable for longer distances. That is, pairs that are more than distance three apart in the co-authorship network have a similar (and minimal) effect on citation patterns. To confirm the interdependence between citation networks and the associated co-authorship network we formulate a null model for the probability of an author $j$ citing author $i$ in Eq. (\ref{eq:pij_ct}) and then using Bayes' formalism (Eq. (6-9)) we explicitly show that the null model is indeed insufficient to explain the citing patterns. There is a significant effect caused by the distance between pairs of authors in the co-authorship network over their citing patterns. The effect being most dominant for immediate co-authors (Fig. \ref{f10}).  
	
	\section{\label{sec:conclusion}Conclusion}
	
	The main contribution of this study lies in a rigorous and a comprehensive analysis that probes the relations between the distances in the co-authorship network and the citation patterns in the citation network. We do this for all possible pairs of researchers from the citation and co-authorship networks constructed. For all pairs of authors, we observe the number of citations exchanged between them as a function of distance in the co-authorship network, as it evolves over time. We find that co-authors dominate the citation patterns in our networks. The remainder of the citations were mostly between pairs that have a short ($d \leq 3$) connected path in the co-authorship network; the average number of citations exchanged decaying with increasing distance. Pairs with distances $d > 3$ have a relatively small contribution to the citations. Disconnected pairs of authors make a small contribution to the citations in the initial years of the network but quickly become almost negligible, as the network grows to be more connected over time. We also highlight the underlying aging effect in mutual citations and collaborations.  
	
    Most of the citations are accounted for within three degrees of separation in our data set. This indicates that authors mostly cite their co-authors and the collaborators of their co-authors. Since we study researchers affiliated to Indian institutes these citations exchanged would be between authors working in similar research topics with a likely connection in the co-authorship network. In short similarity in research topic and affinity towards close collaborators would be the major effects driving the citation patterns in our case. However as some authors (or papers) gain more citations over time the dynamics for the top cited authors (papers) are likely to differ from those of the majority. In such cases, distant or disconnected authors will also have a significant contribution to the total number of citations of the author (or paper). Therefore to measure the impact of a paper with respect to citations received we should have measures that account for this possible bifurcation in citing patterns.       
	
    The main advantage of our data set was in calculating pairwise interactions. Even so, we realize that our data set is not completely comprehensive, as it considers only Indian authors and APS publications, hence our results might show small variations when calculated for larger data sets. However, we believe that this difference should not be considerably large, assuming that pairwise interactions between authors would be similar irrespective of the size of the network. 
    
    We reflect on the opinion that most real-world networks can be viewed as interdependent multi-layer networks, with networks of scientific collaborations as an example of this. This interdependence is critical when studying the dynamics of such networks. Our analysis explicitly shows that connected paths in one network (co-authorship) impact the structure of the other (citation). Dominance of pairs close in distance highlight the importance of an authors' neighborhood in her citing patterns which, in turn, can be used to explain the patterns in flow of ideas and information in the scientific ecosystem. Results from this study can be used in the development of more sophisticated models to investigate the spread of scientific knowledge. We believe that to understand the true nature of research collaboration, it is important to consider both co-authorship and citation networks simultaneously. 
	
	\section{\label{sec:acknowledgement}Acknowledgement}
	
	The authors would like to thank Anand Sengupta, Anirban Dasgupta, Krishna Kanti Dey, Jayesh Choudhary and Amit Reza for their valuable insights and discussions during the period of this study.

\end{document}